\documentclass{article}

\usepackage{microtype}
\usepackage{graphicx}
\usepackage{subcaption}
\usepackage{booktabs} % for professional tables

\usepackage{hyperref}

\usepackage[accepted]{icml2026}

\usepackage{amsmath}
\usepackage{amssymb}
\usepackage{mathtools}
\usepackage{amsthm}
\usepackage{booktabs}

\usepackage[capitalize,noabbrev]{cleveref}

\theoremstyle{plain}

\theoremstyle{definition}

\theoremstyle{remark}

\usepackage[textsize=tiny]{todonotes}

\icmltitlerunning{Multimodal Video-to-Music Recommendation via Semantic Retrieval and Temporal Reranking}

\begin{document}

\twocolumn[
    \icmltitle{Multimodal Video-to-Music Recommendation\\via Semantic Retrieval and Temporal Reranking}
    \icmlsetsymbol{equal}{*}
    \begin{icmlauthorlist}
    \icmlauthor{Seungheon Doh}{kaist,equal}
    \icmlauthor{Minhee Lee}{kaist}
    \icmlauthor{Sangmoon Lee}{gaudio}
    \icmlauthor{Ben Sangbae Chon}{gaudio}
    \icmlauthor{Juhan Nam}{kaist}
    \end{icmlauthorlist}
    \begin{icmlauthorlist}
    \text{\color{magenta}{https://seungheondoh.github.io/video-to-music-demo}}
    \end{icmlauthorlist}
    \icmlaffiliation{kaist}{Graduate School of Culture Technology, KAIST, South Korea}
    \icmlaffiliation{gaudio}{Gaudio Lab, Inc, South Korea. $^{*}$Work completed while Seungheon was visiting Gaudio Lab}
    \icmlcorrespondingauthor{Seungheon Doh}{seungheon.doh@gmail.com}
    \icmlkeywords{Music Recommendation, Video-to-Music Retrieval}
    \vskip 0.3in
]

% this must go after the closing bracket ] following \twocolumn[ ...

% This command actually creates the footnote in the first column listing the
% affiliations and the copyright notice. The command takes one argument, which
% is text to display at the start of the footnote. The \icmlEqualContribution
% command is standard text for equal contribution. Remove it (just {}) if you
% do not need this facility.

% Use ONE of the following lines. DO NOT remove the command.
% If you have no special notice, KEEP empty braces:
\printAffiliationsAndNotice{}  % no special notice (required even if empty)
% Or, if applicable, use the standard equal contribution text:
% \printAffiliationsAndNotice{\icmlEqualContribution}

\begin{abstract}
We present \textbf{VTMR}, a two-stage framework for \textbf{V}ideo-\textbf{T}o-\textbf{M}usic \textbf{R}ecommendation. In Stage~1, VTMR aligns comprehensive video and music signals in a joint audio-visual-text representation space and efficiently retrieves semantically compatible candidates using coarse global embeddings. In Stage~2, it reranks the retrieved candidates by attending to the temporal sequences of both video and music, thereby capturing fine-grained temporal correspondence. Evaluated on the video-to-music recommendation task, the multimodal retrieval stage improves R@10 from 14.2 to 15.9 and Median Rank from 75 to 58 over the strongest baseline; the temporal reranker further boosts R@10 to 18.3 and Median Rank to 46, demonstrating complementary gains from richer query encoding and temporal alignment. A human preference study confirms that VTMR is on par with a commercial baseline in overall preference, while outperforming a generative baseline in music quality.
\end{abstract}

\vspace{-5mm}
\section{Introduction}
Music is a cornerstone of compelling video content, shaping emotional tone, narrative pacing, and audience engagement across domains from cinematic productions to short-form social media. Yet selecting appropriate background music is far from trivial: it requires not only an understanding of high-level semantic compatibility, such as mood and genre, but also a fine-grained understanding of the temporal correspondence between evolving visual dynamics and musical elements. Identifying music that satisfies both criteria is time-consuming and often yields suboptimal results.

Automatic video-to-music (V2M) recommendation has therefore attracted growing interest as a cross-modal music retrieval task~\citep{li2019query, huang2022mulan, doh2023toward, wu2025clamp}, aiming to learn joint embedding spaces between visual content and music~\citep{suris2022time, pretet2023video, wilkins2023bridging, stewart2025semi}. Despite steady progress, current frameworks leave substantial room for improvement across two major dimensions. First, while general-purpose multimodal models~\citep{guzhov2022audioclip, girdhar2023imagebind, zhu2024languagebind} have demonstrated the power of joint audio-visual-text representations, established V2M methods still rely almost exclusively on raw visual features. Even though some previous works~\citep{mckee2023language} have utilized both visual and textual modalities for V2M recommendation, existing architectures have yet to fully exploit the multimodal richness of both video and music signals.

Second, all existing V2M methods reduce retrieval to a single global embedding similarity score, compressing entire video and music streams into static vectors. While global embeddings are effective for capturing coarse semantic compatibility, such as overall mood or genre, this single-vector bottleneck fundamentally cannot represent how localized musical events align with specific video moments. The temporal dimension is collapsed by design, making it impossible to distinguish music that is globally \emph{compatible} from music that is temporally \emph{aligned} with specific video moments.

\begin{figure*}[!t]
\centering
\includegraphics[width=\linewidth]{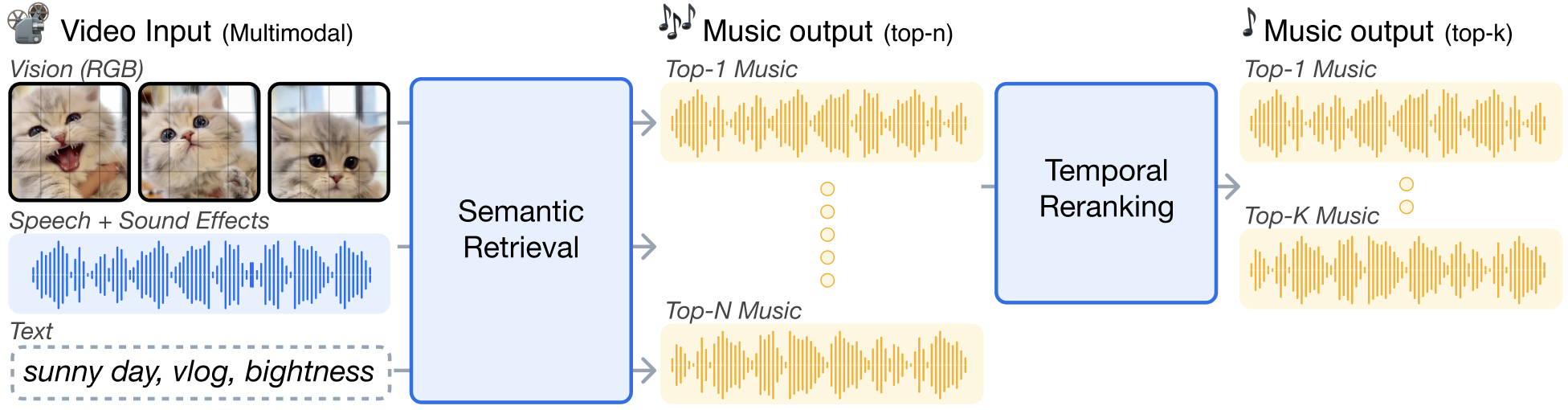}
\vspace{-1mm}
\caption{\textbf{Overview of VTMR.} Stage~1 projects multimodal video and music signals into a shared embedding space and retrieves the top-$N$ candidates via global similarity. Stage~2 reranks each candidate by attending over the temporal sequences of the video-music pair.}
\vspace{-3mm}
\label{fig:overview}
\end{figure*}

To address both limitations, we propose \textbf{VTMR}, a two-stage framework capable of capturing both unified multimodal semantics and temporal dynamics (see Figure~\ref{fig:overview}). \textit{Stage~1 (Semantic Retrieval)} addresses the first limitation by projecting comprehensive multimodal video signals, music signals, and LLM-generated text descriptions into a tightly aligned shared embedding space, enabling efficient large-scale retrieval of the top-$N$ candidates.
\textit{Stage~2 (Temporal Reranking)} resolves the second limitation by deploying a fine-grained cross-encoder. It directly attends over the dense, unpooled temporal sequences of both the video and the retrieved music candidates, capturing the intricate cross-modal dynamics that global embeddings inevitably lose.

\section{Methods}
Let $\mathcal{M} = \{m_1, \dots, m_n\}$ be a music corpus of $n$ tracks. Given a query video $v$, the goal is to retrieve and rank tracks from $\mathcal{M}$ by relevance. We represent $v$ as a multimodal tuple $(x^v, x^a, x^t)$, where $x^v$ denotes raw RGB frames, $x^a$ the non-musical audio (speech, sound effects), and $x^t$ an LLM-generated scene description. Symmetrically, each music track $m \in \mathcal{M}$ is defined as a multimodal composition $(x^m, x^{mt})$, where $x^m$ represents the acoustic music audio stream and $x^{mt}$ denotes the LLM-generated music caption. Additionally, $x^{meta}$ denotes raw video metadata (e.g., title, channel, category information).

\subsection{Multimodal Media Encoding}
\label{sec:encoding}

We use \text{Perception Encoder Audiovisual} 
(PE\textsubscript{AV-base})~\citep{vyas2025pushing} as a frozen backbone to encode all modalities. PE\textsubscript{AV-base} is trained on $\mathcal{O}(100\text{M})$ synthetic audiovisual pairs via multiple cross-modal contrastive objectives.

\vspace{-1mm}\paragraph{Video Encoding.}
A video $v = (x^v, x^a)$ contains two complementary streams. The {visual stream} $x^v$ is encoded by the PE-L frame encoder into a dense temporal sequence $\mathbf{e}^v = {PE_{frame}}(x^v) \in \mathbb{R}^{L_v \times C}$. The \emph{non-musical audio stream} $x^a$ (speech, sound effects) is tokenized by DAC-VAE at 25\,Hz and contextualized by Transformer Layers with RoPE to yield $\mathbf{e}^a = {PE_{audio}}(x^a) \in \mathbb{R}^{L_a \times C}$. The two streams are temporally aligned and fused by a shallow Transformer into a joint audiovisual \texttt{[CLS]} token $\mathbf{h}^{av} = {PE_{fuse}}(\mathbf{e}^v,\, \mathbf{e}^a)_{\texttt{[CLS]}} \in \mathbb{R}^{C}$, which compactly represents \emph{what is seen and heard} in the video.

\vspace{-1mm}\paragraph{Music Encoding.}
A music track $m = (x^m)$ is encoded by the same PE\textsubscript{AV} audio encoder into a temporal sequence $\mathbf{e}^m = {PE_{audio}}(x^m) \in \mathbb{R}^{L_m \times C}$. 
A global music embedding $\mathbf{h}^m \in \mathbb{R}^C$ is obtained by taking the \texttt{[CLS]} token.

\vspace{-1mm}\paragraph{Text Encoding.}
The video scene description $x^t$, music caption $x^{mt}$, and video metadata $x^{meta}$ are encoded by the PE\textsubscript{AV} text encoder (ModernBERT), yielding global representations $\mathbf{h}^{t} = {PE_{text}}(x^t) \in \mathbb{R}^{C}$, $\mathbf{h}^{mt} = {PE_{text}}(x^{mt}) \in \mathbb{R}^{C}$, and $\mathbf{h}^{meta} = {PE_{text}}(x^{meta}) \in \mathbb{R}^{C}$, respectively. The global representations~$(\mathbf{h})$ feed the Stage~1 semantic retrieval, while the temporal sequences~$(\mathbf{e})$ feed the Stage~2 temporal reranker.

\subsection{Semantic Retrieval}
\label{sec:retrieval}

\vspace{1mm}\paragraph{Retrieval Modules.} We map the global representations extracted from the pre-trained encoders (Section~\ref{sec:encoding}) into a shared $d$-dimensional embedding space ($d{=}1024$) to facilitate efficient large-scale retrieval. Formally, we define the projection networks $f_v$, $f_m$, and $f_t$ for the video, music, and text modalities, respectively, where each head consists of a LayerNorm followed by a bias-free 3-layer MLP with intermediate GELU activations. The resulting joint embeddings are formalized as the set $\mathcal{Z} = \{\mathbf{z}^{v}, \mathbf{z}^{m}, \mathbf{z}^{vt}, \mathbf{z}^{mt}, \mathbf{z}^{meta}\}$, where the isolated projected vectors are mapped via $\mathbf{z}^{v} = f_v(\mathbf{h}^{av})$, $\mathbf{z}^{m} = f_m(\mathbf{h}^{m})$, $\mathbf{z}^{vt} = f_t(\mathbf{h}^{t})$, $\mathbf{z}^{mt} = f_t(\mathbf{h}^{mt})$, and $\mathbf{z}^{meta} = f_t(\mathbf{h}^{meta})$.

\vspace{-1mm}\paragraph{Modality Fusion and Dropout.} To ensure the model remains robust to missing modalities at inference time (e.g., querying with videos that lack text descriptions), we employ a dynamic modality fusion and dropout strategy during training~\citep{bang2025pianobind,choi2021listen}. For a given batch, with a probability of $0.5$, we fuse the parallel representations within $\mathcal{Z}$ by mean-pooling the projected media and text embeddings, dynamically updating the active video and music keys via $\mathbf{z}^{v} \leftarrow \frac{1}{2}(\mathbf{z}^{v} + \mathbf{z}^{vt})$ and $\mathbf{z}^{m} \leftarrow \frac{1}{2}(\mathbf{z}^{m} + \mathbf{z}^{mt})$. Otherwise, we drop the text streams entirely and rely solely on the isolated audiovisual ($\mathbf{z}^{v}$) or acoustic ($\mathbf{z}^{m}$) vectors.

\vspace{-1mm}\paragraph{Loss Function.} For training, the semantic retrieval module is optimized using a multi-pair contrastive framework based on the SigLIP loss~\citep{tschannen2025siglip}, where pairwise sigmoid log-likelihoods are computed across all $\binom{5}{2} = 10$ unique modality combinations derived from the projected embedding set $\mathcal{Z}$. 

% Formally, the loss between any two modality matrices $\mathbf{Z}_a, \mathbf{Z}_b \in \mathbb{R}^{B \times d}$ is measured as:
% $$
% \begin{aligned}
% \mathcal{L}_{\text{SigLIP}}(\mathbf{Z}_a, \mathbf{Z}_b) = 
% -\frac{1}{B} \sum_{i=1}^{B} \log \sigma (t \cdot \mathbf{z}_{a,i}^\top \mathbf{z}_{b,i} + b) \\
% -\frac{1}{B(B-1)} \sum_{i=1}^{B} \sum_{j \neq i} \log \left(1 - \sigma (t \cdot \mathbf{z}_{a,i}^\top \mathbf{z}_{b,j} + b)\right)
% \end{aligned}
% $$
% which is driven by a learnable logit scale $t$ (initialized to $\ln(10)$) and bias $b$ (initialized to $-10.0$). The total training objective is formulated as the joint average $\mathcal{L}_{\text{total}} = \frac{1}{10} \sum_{(a,b)} \lambda_{a,b} \mathcal{L}_{\text{SigLIP}}(\mathbf{Z}_a, \mathbf{Z}_b)$, where this dense cross-modal supervision forces all representations into a tightly aligned semantic space. 

\subsection{Temporal Reranking}
\label{sec:reranking}

\paragraph{Architecture.} Given the top-$N$ candidates retrieved in Stage~1, the temporal reranker scores each pair $(v, m)$ using a \emph{cross-encoder} architecture~\citep{humeau2019poly, zhang2023baai}. We extract the audio-visual frame sequence $\mathbf{x}^{{av}}$ and the music acoustic feature sequence $\mathbf{x}^{{m}}$ using PE\textsubscript{AV-base}~\citep{vyas2025pushing}. To accommodate variable-length inputs, both sequences are resampled along the temporal dimension to a fixed length ($T_{\text{target}} = 64$), using \emph{window averaging} for downsampling and \emph{linear interpolation} for upsampling. The resampled sequences are then linearly projected into a shared $d_j$-dimensional space, yielding aligned representations $\mathbf{e}^{{av}}, \mathbf{e}^{{m}} \in \mathbb{R}^{T_{\text{target}} \times d_j}$. To distinguish the two modalities while allowing their temporal representations to interact, we add learnable modality-type embeddings $\mathbf{t}^{{av}}, \mathbf{t}^{{m}} \in \mathbb{R}^{d_j}$ and construct the concatenated embedding sequence $\mathbf{e}^{v+m} = [\,\texttt{[CLS]};\; \mathbf{e}^{{av}} + \mathbf{t}^{{av}};\; \texttt{[SEP]};\; \mathbf{e}^{{m}} + \mathbf{t}^{{m}}\,]$. A 4-layer Transformer encoder with 8 attention heads then applies self-attention over the entire concatenated sequence. Finally, the contextualized \texttt{[CLS]} representation is passed through a two-layer GELU MLP head to produce a scalar classifier logit $s(v,m)$.

\vspace{-2mm}\paragraph{Loss Functions.} During training, the reranker is optimized using a hybrid objective. For each query video $v$, its paired music track $m^+$ serves as the positive, while a negative track $m^-$ is sampled from the other music tracks in the batch based on their semantic retrieval scores~(Section~\ref{sec:retrieval}). We compute the corresponding logits as $s^+=s(v,m^+)$ and $s^-=s(v,m^-)$. To encourage both pointwise discrimination and relative ordering, we combine binary cross-entropy (BCE) with a margin-based ranking loss: $\mathcal{L}_{\text{rerank}}=-\log\sigma(s^+)-\log(1-\sigma(s^-))+\max(0,\gamma-(s^+-s^-))$, where $\sigma(\cdot)$ denotes the sigmoid function and $\gamma=0.2$ is the ranking margin.

\vspace{-2mm}
\section{Dataset}
\begin{figure}[!t]
\centering
\includegraphics[width=\linewidth]{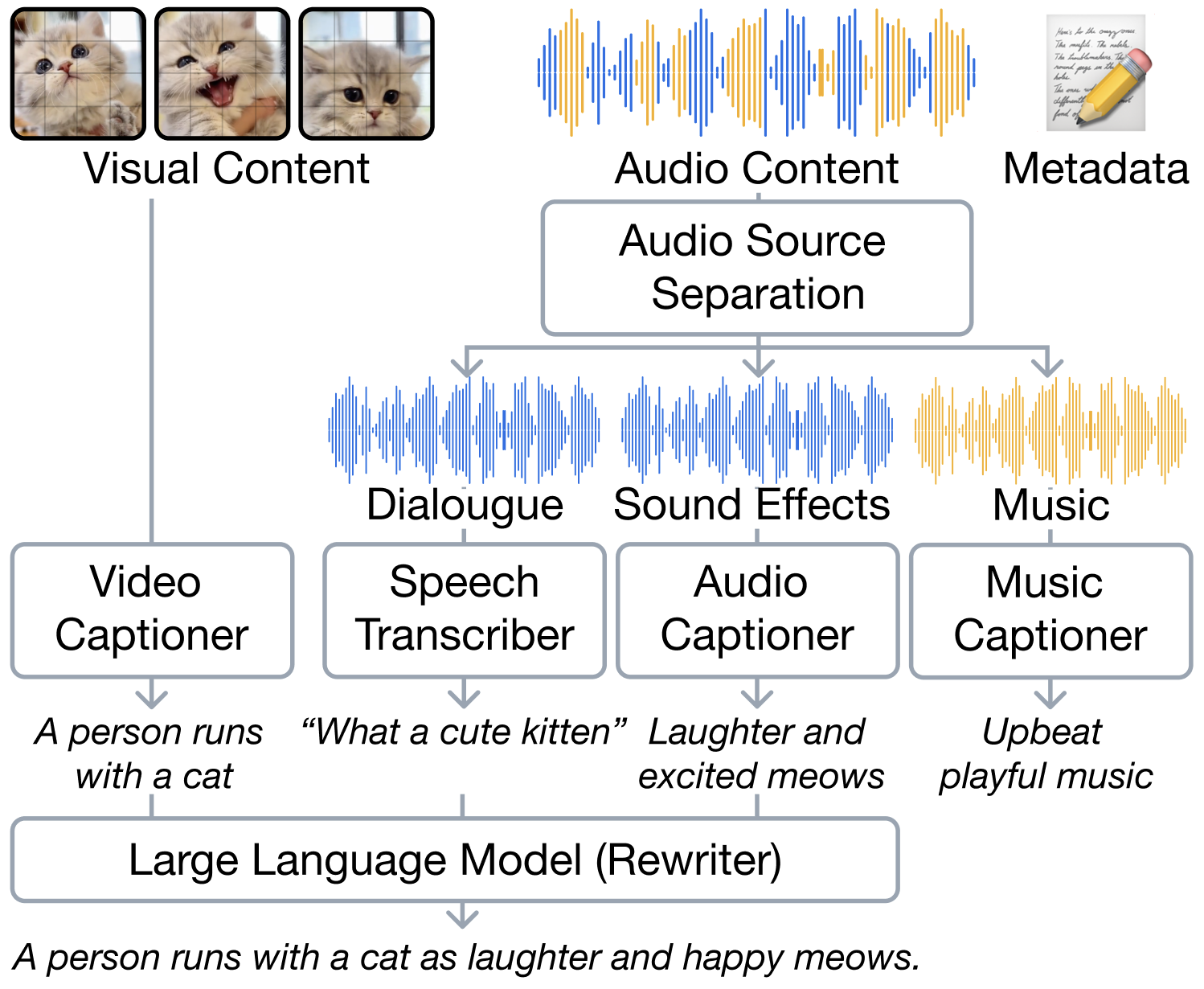}
\caption{Overview of the preprocessing pipeline.}
\vspace{-5mm}
\label{fig:preprocessing}
\end{figure}

To build a robust framework for multimodal video-to-music recommendation, we construct our dataset through a three-stage pipeline: collecting videos from diverse source datasets, filtering them to retain salient and high-quality background music, and generating pseudo annotations from visual and separated audio streams (illustrated in Figure~\ref{fig:preprocessing}).

\vspace{-2mm}\paragraph{Source Datasets.} To build a diverse media dataset, we establish three collection criteria. First, the videos should preserve their original mixed audio, containing real-world combinations of speech, sound effects, and background music. Second, background music should be present for at least 50\% of each video's duration, ensuring a sufficiently strong video--music association. Third, the dataset should cover diverse content sources and production contexts, ranging from professionally produced broadcast programs to user-generated YouTube videos. Our data are drawn from three distinct sources: (1) \textsc{MMtrail-2M}~\citep{chi2024mmtrail}, a large-scale public dataset of YouTube videos; (2) \textsc{MovieLens-Content}~\citep{lee2017large}, a dataset of 16K movie trailers produced by professional filmmakers; and (3) an internal dataset comprising broadcast content.

\vspace{-2mm}\paragraph{Dataset Filtering.} To ensure the fidelity and semantic relevance of the target background music, we apply a rigorous filtering process. First, we filter out non-music video clips based on music logits derived from a pre-trained music detection model~\citep{kong2020panns}, discarding samples with a music logit score of 0.7 or lower. Furthermore, we exclude video sequences exceeding 2 minutes to maintain dense temporal alignment and optimize computational efficiency. Finally, to guarantee music quality, we utilize an aesthetic prediction model~\citep{tjandra2025meta}, removing music tracks with a content enjoyment score below 5.5.

\vspace{-2mm}\paragraph{Pseudo Annotation.} To enable scalable content descriptions, we leverage pseudo annotations~\citep{doh2023lp,doh2024enriching,doh2025talkplay}. Since the original audio contains overlapping speech, sound effects, and background music, we employ an audio source separation model~\citep{gaudiolab_source_separation} to disentangle the mixed audio into three independent streams: speech, sound effects, and music. Following this audio separation, we annotate semantic descriptions by employing a suite of multimodal Large Language Models. Specifically, we extract visual captions using \text{Qwen3-VL-2B-Instruct}~\citep{bai2025qwen3}, transcribe spoken dialogue using \text{Qwen3-ASR-1.7B}~\citep{shi2026qwen3}, capture ambient sound contexts via \text{AudioFlamingo-3}~\citep{ghosh2026audio}, and describe musical attributes using \text{MusicFlamingo}~\citep{ghosh2025music}. Finally, for audio-visual description, we aggregate vision, speech, and sound-effect descriptions with an \text{LLM rewriter} powered by \text{Qwen3-4B-Instruct}~\citep{yang2025qwen3}.

\section{Experiments}

\begin{table}[t]
\centering
\caption{Video-to-music recommendation performance.}
\vspace{-2mm}
\label{tab:quantitative}
\resizebox{\columnwidth}{!}{% 
\begin{tabular}{lcc}
\toprule
Model & R@10 ($\uparrow$) & MedR ($\downarrow$) \\
\midrule
\multicolumn{3}{l}{\textcolor{gray}{\textit{Baseline Methods}}} \\
Random                               & 4.4                       & 141                          \\
AudioCLIP~\citep{guzhov2022audioclip} & 6.1                       & 116                          \\
Wav2CLIP~\citep{wu2022wav2clip}     & 6.7                       & 117                          \\
ImageBind~\citep{girdhar2023imagebind} & 14.2                      & 75                           \\
LanguageBind~\citep{zhu2024languagebind} & 10.2                      & 84                           \\
PE\textsubscript{AV-base}~\citep{vyas2025pushing}       & 11.5                      & 87                           \\
\midrule
\multicolumn{3}{l}{\textcolor{gray}{\textit{Proposed Methods}}} \\
VTMR (w/o Reranker)                  & \underline{15.9}          & \underline{58}               \\
VTMR (w/ Reranker)                   & \textbf{18.3}             & \textbf{46}                  \\
\bottomrule
\end{tabular}%              
}
\vspace{-3mm}
\end{table}

\paragraph{Quantitative Evaluation.} To evaluate our video-to-music recommendation system, we assess cross-modal retrieval performance on the human-verified VidMuse benchmark~\citep{tian2025vidmuse}. We re-crawled the dataset to incorporate all multimodal signals extracted via our preprocessing pipeline. At inference, the video query vector is obtained by averaging the audiovisual and text embeddings; the music vector averages its audio and text embeddings. We use re-ranker with-top 40 retrieval sample. We report Recall@10 (R@10) and Median Rank (MedR) as primary metrics, benchmarking against AudioCLIP~\citep{guzhov2022audioclip}, Wav2CLIP~\citep{wu2022wav2clip}, ImageBind~\citep{girdhar2023imagebind}, LanguageBind~\citep{zhu2024languagebind}, and PE\textsubscript{AV-base}~\citep{vyas2025pushing}.

Table~\ref{tab:quantitative} summarizes the quantitative evaluation results. Our proposed multimodal fusion retrieval module, VTMR (w/o Reranker), consistently outperforms all existing baseline models. Notably, it yields a substantial margin of improvement over the strongest baseline, ImageBind, increasing the R@10 from 14.2 to 15.9 and reducing the Median Rank from 75 to 58. Furthermore, integrating the Stage 2 temporal reranker—VTMR (w/ Reranker)—delivers a powerful performance boost, achieving the highest R@10 of 18.3 and reducing the MedR further to 46. This remarkable improvement underscores the critical importance of modeling fine-grained, unpooled temporal sequences via cross-encoder to align cross-modal dynamics effectively.

\begin{figure}[!t]
\centering
\includegraphics[width=\linewidth]{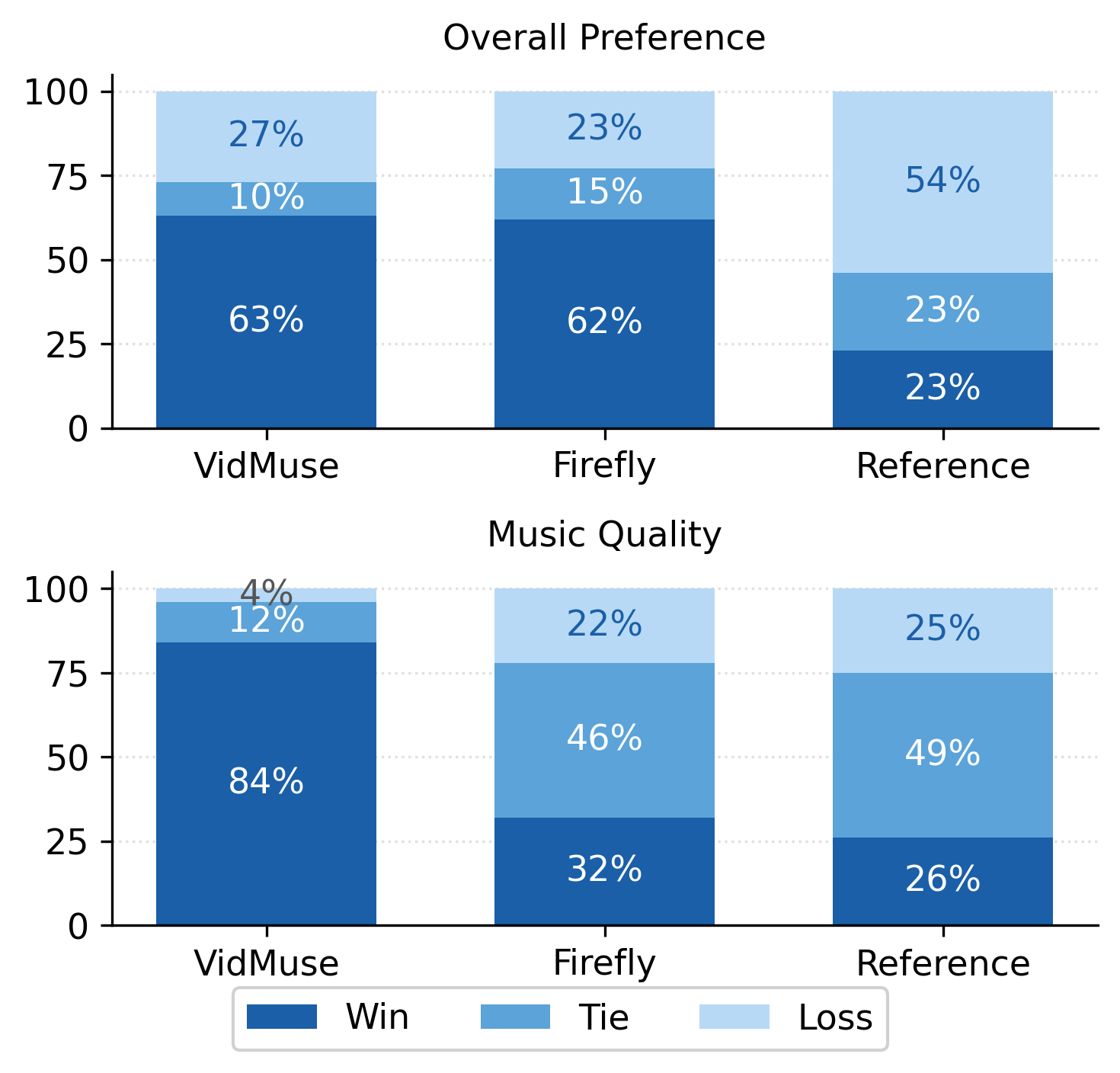}
\vspace{-7mm}
\caption{A-vs-B Subjective Evaluation Results.(VTMR vs.)}
\vspace{-7mm}
\label{fig:humaneval}
\end{figure}

\vspace{-2mm}\paragraph{Qualitative Evaluation.} To evaluate the subjective quality of the video-to-music system, we conducted a human evaluation using an A-vs-B preference test. We selected 20 diverse video clips as evaluation stimuli: 10 professional broadcast clips featuring original scores curated by music directors, and 10 background-music-free YouTube clips. To ensure high-quality retrieval candidates, our model retrieves from an external music database of 500K professionally produced tracks, providing broad coverage across genres, moods, and instrumentation. Our proposed framework was benchmarked against three distinct configurations: (1) \emph{Human-Reference} (upperbound), (2) \emph{Adobe Firefly} (commercial baseline), and (3) \emph{VidMuse}~\citep{tian2025vidmuse} (generative model baseline).

Human annotators evaluated the paired samples based on two primary perceptual dimensions: \textit{Overall Preference} and \textit{Music Quality}. A total of 30 expert evaluators—comprising academic music researchers and audio industry professionals—participated in the user study, yielding a robust dataset of 1,200 pairwise preference annotations.

As shown in Figure~\ref{fig:humaneval}, the win+tie rates reveal two complementary trends. In terms of \textit{Overall Preference}, our model is competitive with both the commercial baseline Adobe Firefly (77\%) and the generative baseline VidMuse (73\%), while falling short of professional human reference curation (46\%), which remains a strong upper bound. In terms of \textit{Music Quality}, our retrieval-based model substantially outperforms the generative model VidMuse (96\% win+tie vs.\ 4\% loss), demonstrating that retrieving from a curated music corpus yields consistently higher intrinsic audio quality than end-to-end generation. Taken together, these results suggest that our framework reaches the practical quality range of commercial tools while avoiding music quality degradation.

\section{Conclusion}
We presented VTMR, a two-stage framework for video-to-music recommendation that combines multimodal semantic correspondence with fine-grained temporal matching. Its key contribution is a retrieve-and-rerank architecture that first identifies semantically compatible music using visual, audio, and textual cues and then captures temporal correspondence between video and music. Future work will explore end-to-end training of the retrieve-and-rerank pipeline.

% In the unusual situation where you want a paper to appear in the
% references without citing it in the main text, use \nocite
% \nocite{langley00}

\bibliography{references}

@article{choi2021listen,
  title={Listen, read, and identify: Multimodal singing language identification of music},
  author={Choi, Keunwoo and Wang, Yuxuan},
  journal={arXiv preprint arXiv:2103.01893},
  year={2021}
}

@article{bang2025pianobind,
  title={PianoBind: A Multimodal Joint Embedding Model for Pop-piano Music},
  author={Bang, Hayeon and Choi, Eunjin and Doh, Seungheon and Nam, Juhan},
  journal={arXiv preprint arXiv:2509.04215},
  year={2025}
}

@inproceedings{li2019query,
  title={Query by Video: Cross-modal Music Retrieval.},
  author={Li, Bochen and Kumar, Aparna},
  booktitle={ISMIR},
  pages={604--611},
  year={2019}
}

@article{doh2025talkplay,
  title={TALKPLAY: Multimodal Music Recommendation with Large Language Models},
  author={Doh, Seungheon and Choi, Keunwoo and Nam, Juhan},
  journal={arXiv preprint arXiv:2502.13713},
  year={2025}
}

@article{doh2023lp,
  title={Lp-musiccaps: Llm-based pseudo music captioning},
  author={Doh, SeungHeon and Choi, Keunwoo and Lee, Jongpil and Nam, Juhan},
  journal={arXiv preprint arXiv:2307.16372},
  year={2023}
}

@article{huang2022mulan,
  title={Mulan: A joint embedding of music audio and natural language},
  author={Huang, Qingqing and Jansen, Aren and Lee, Joonseok and Ganti, Ravi and Li, Judith Yue and Ellis, Daniel PW},
  journal={International Society for Music Information Retrieval (ISMIR)},
  year={2022}
}

@inproceedings{doh2023toward,
  title={Toward universal text-to-music retrieval},
  author={Doh, SeungHeon and Won, Minz and Choi, Keunwoo and Nam, Juhan},
  booktitle={ICASSP 2023-2023 IEEE International Conference on Acoustics, Speech and Signal Processing (ICASSP)},
  year={2023},  
}

@inproceedings{doh2024enriching,
  title={Enriching Music Descriptions with A Finetuned-{LLM} and Metadata for Text-to-Music Retrieval},
  author={Doh, SeungHeon and Lee, Minhee and Jeong, Dasaem and Nam, Juhan},
  booktitle={ICASSP 2024-2024 IEEE International Conference on Acoustics, Speech and Signal Processing (ICASSP)},
  year={2024},
}

@article{wu2025clamp,
  title={Clamp 3: Universal music information retrieval across unaligned modalities and unseen languages},
  author={Wu, Shangda and Guo, Zhancheng and Yuan, Ruibin and Jiang, Junyan and Doh, Seungheon and Xia, Gus and Nam, Juhan and Li, Xiaobing and Yu, Feng and Sun, Maosong},
  journal={arXiv preprint arXiv:2502.10362},
  year={2025}
}

@inproceedings{wilkins2023bridging,
  title={Bridging high-quality audio and video via language for sound effects retrieval from visual queries},
  author={Wilkins, Julia and Salamon, Justin and Fuentes, Magdalena and Bello, Juan Pablo and Nieto, Oriol},
  booktitle={2023 IEEE Workshop on Applications of Signal Processing to Audio and Acoustics (WASPAA)},
  pages={1--5},
  year={2023},
  organization={IEEE}
}

@inproceedings{tian2025vidmuse,
  title={Vidmuse: A simple video-to-music generation framework with long-short-term modeling},
  author={Tian, Zeyue and Liu, Zhaoyang and Yuan, Ruibin and Pan, Jiahao and Liu, Qifeng and Tan, Xu and Chen, Qifeng and Xue, Wei and Guo, Yike},
  booktitle={Proceedings of the Computer Vision and Pattern Recognition Conference},
  pages={18782--18793},
  year={2025}
}

@inproceedings{zhu2024languagebind,
  title={Languagebind: Extending video-language pretraining to n-modality by language-based semantic alignment},
  author={Zhu, Bin and Lin, Bin and Ning, Munan and Yan, Yang and Cui, Jiaxi and HongFa, WANG and Pang, Yatian and Jiang, Wenhao and Zhang, Junwu and Li, Zongwei and others},
  booktitle={International Conference on Learning Representations},
  volume={2024},
  pages={9588--9608},
  year={2024}
}

@inproceedings{girdhar2023imagebind,
  title={Imagebind: One embedding space to bind them all},
  author={Girdhar, Rohit and El-Nouby, Alaaeldin and Liu, Zhuang and Singh, Mannat and Alwala, Kalyan Vasudev and Joulin, Armand and Misra, Ishan},
  booktitle={Proceedings of the IEEE/CVF conference on computer vision and pattern recognition},
  pages={15180--15190},
  year={2023}
}

@inproceedings{wu2022wav2clip,
  title={Wav2clip: Learning robust audio representations from clip},
  author={Wu, Ho-Hsiang and Seetharaman, Prem and Kumar, Kundan and Bello, Juan Pablo},
  booktitle={ICASSP 2022-2022 IEEE International Conference on Acoustics, Speech and Signal Processing (ICASSP)},
  pages={4563--4567},
  year={2022},
  organization={IEEE}
}

@inproceedings{guzhov2022audioclip,
  title={Audioclip: Extending clip to image, text and audio},
  author={Guzhov, Andrey and Raue, Federico and Hees, J{\"o}rn and Dengel, Andreas},
  booktitle={ICASSP 2022-2022 IEEE International Conference on Acoustics, Speech and Signal Processing (ICASSP)},
  pages={976--980},
  year={2022},
  organization={IEEE}
}

@article{tjandra2025meta,
  title={Meta audiobox aesthetics: Unified automatic quality assessment for speech, music, and sound},
  author={Tjandra, Andros and Wu, Yi-Chiao and Guo, Baishan and Hoffman, John and Ellis, Brian and Vyas, Apoorv and Shi, Bowen and Chen, Sanyuan and Le, Matt and Zacharov, Nick and others},
  journal={arXiv preprint arXiv:2502.05139},
  year={2025}
}

@misc{gaudiolab_source_separation,
  author       = {{Gaudio Lab Inc.}},
  title        = {Gaudio source separation: Dialogue-Music-Sound-Effects Seperator},
  year         = {2026},
  howpublished = {\url{https://www.gaudiolab.com/technology/source-separation}},
  note         = {Accessed: 2026-05-26}
}

@article{yang2025qwen3,
  title={Qwen3 technical report},
  author={Yang, An and Li, Anfeng and Yang, Baosong and Zhang, Beichen and Hui, Binyuan and Zheng, Bo and Yu, Bowen and Gao, Chang and Huang, Chengen and Lv, Chenxu and others},
  journal={arXiv preprint arXiv:2505.09388},
  year={2025}
}

@article{ghosh2025music,
  title={Music flamingo: Scaling music understanding in audio language models},
  author={Ghosh, Sreyan and Goel, Arushi and Koroshinadze, Lasha and Lee, Sang-gil and Kong, Zhifeng and Santos, Joao Felipe and Duraiswami, Ramani and Manocha, Dinesh and Ping, Wei and Shoeybi, Mohammad and others},
  journal={arXiv preprint arXiv:2511.10289},
  year={2025}
}

@article{ghosh2026audio,
  title={Audio flamingo 3: Advancing audio intelligence with fully open large audio language models},
  author={Ghosh, Sreyan and Goel, Arushi and Kim, Jaehyeon and Kumar, Sonal and Kong, Zhifeng and Lee, Sang-gil and Yang, Chao-Han and Duraiswami, Ramani and Manocha, Dinesh and Valle, Rafael and others},
  journal={Advances in Neural Information Processing Systems},
  volume={38},
  pages={41819--41886},
  year={2026}
}

@article{shi2026qwen3,
  title={Qwen3-ASR Technical Report},
  author={Shi, Xian and Wang, Xiong and Guo, Zhifang and Wang, Yongqi and Zhang, Pei and Zhang, Xinyu and Guo, Zishan and Hao, Hongkun and Xi, Yu and Yang, Baosong and others},
  journal={arXiv preprint arXiv:2601.21337},
  year={2026}
}

@article{bai2025qwen3,
  title={Qwen3-vl technical report},
  author={Bai, Shuai and Cai, Yuxuan and Chen, Ruizhe and Chen, Keqin and Chen, Xionghui and Cheng, Zesen and Deng, Lianghao and Ding, Wei and Gao, Chang and Ge, Chunjiang and others},
  journal={arXiv preprint arXiv:2511.21631},
  year={2025}
}

@article{kong2020panns,
  title={Panns: Large-scale pretrained audio neural networks for audio pattern recognition},
  author={Kong, Qiuqiang and Cao, Yin and Iqbal, Turab and Wang, Yuxuan and Wang, Wenwu and Plumbley, Mark D},
  journal={IEEE/ACM Transactions on Audio, Speech, and Language Processing},
  volume={28},
  pages={2880--2894},
  year={2020},
  publisher={IEEE}
}

@inproceedings{lee2017large,
  title={Large-scale content-only video recommendation},
  author={Lee, Joonseok and Abu-El-Haija, Sami},
  booktitle={Proceedings of the IEEE International Conference on Computer Vision Workshops},
  pages={987--995},
  year={2017}
}

@article{chi2024mmtrail,
  title={Mmtrail: A multimodal trailer video dataset with language and music descriptions},
  author={Chi, Xiaowei and Wang, Yatian and Cheng, Aosong and Fang, Pengjun and Tian, Zeyue and He, Yingqing and Liu, Zhaoyang and Qi, Xingqun and Pan, Jiahao and Zhang, Rongyu and others},
  journal={arXiv preprint arXiv:2407.20962},
  year={2024}
}

@article{humeau2019poly,
  title={Poly-encoders: Transformer architectures and pre-training strategies for fast and accurate multi-sentence scoring},
  author={Humeau, Samuel and Shuster, Kurt and Lachaux, Marie-Anne and Weston, Jason},
  journal={arXiv preprint arXiv:1905.01969},
  year={2019}
}

@article{zhang2023baai,
  title={BAAI General Embedding: A Comprehensive and Efficient Embedding Model for Retrieval and Retrieval-Augmented Generation},
  author={Zhang, Shitao and Liu, Zheng and Xiao, Shitao and Jiang, Jian-Yun},
  journal={arXiv preprint arXiv:2309.07597},
  year={2023}
}

@article{vyas2025pushing,
  title={Pushing the Frontier of Audiovisual Perception with Large-Scale Multimodal Correspondence Learning},
  author={Vyas, Apoorv and Chang, Heng-Jui and Yang, Cheng-Fu and Huang, Po-Yao and Gao, Luya and Richter, Julius and Chen, Sanyuan and Le, Matt and Doll{\'a}r, Piotr and Feichtenhofer, Christoph and others},
  journal={arXiv preprint arXiv:2512.19687},
  year={2025}
}

@article{tschannen2025siglip,
  title={Siglip 2: Multilingual vision-language encoders with improved semantic understanding, localization, and dense features},
  author={Tschannen, Michael and Gritsenko, Alexey and Wang, Xiao and Naeem, Muhammad Ferjad and Alabdulmohsin, Ibrahim and Parthasarathy, Nikhil and Evans, Talfan and Beyer, Lucas and Xia, Ye and Mustafa, Basil and others},
  journal={arXiv preprint arXiv:2502.14786},
  year={2025}
}

@inproceedings{suris2022time,
  author    = {D{\'\i}dac Sur{\'\i}s and Carl Vondrick and Bryan Russell and Justin Salamon},
  title     = {It's Time for Artistic Correspondence in Music and Video},
  booktitle = {Proceedings of the IEEE/CVF Conference on Computer Vision and Pattern Recognition (CVPR)},
  year      = {2022},
}

@inproceedings{mckee2023language,
  author    = {Daniel McKee and Justin Salamon and Josef Sivic and Bryan Russell},
  title     = {Language-Guided Music Recommendation for Video via Prompt Analogies},
  booktitle = {Proceedings of the IEEE/CVF Conference on Computer Vision and Pattern Recognition (CVPR)},
  year      = {2023},
}

@article{pretet2023video,
  author    = {Laure Pr{\'e}tet and Ga{\"e}l Richard and Cl{\'e}ment Souchier and Geoffroy Peeters},
  title     = {Video-to-Music Recommendation using Temporal Alignment of Segments},
  journal   = {IEEE Transactions on Multimedia},
  year      = {2023},
}

@inproceedings{stewart2025semi,
  author    = {Shanti Stewart and Gouthaman KV and Andrea Fanelli and Lie Lu},
  title     = {Semi-Supervised Contrastive Learning for Controllable Video-to-Music Retrieval},
  booktitle = {Proceedings of the IEEE International Conference on Acoustics, Speech and Signal Processing (ICASSP)},
  year      = {2025},
}
\bibliographystyle{icml2026}

%%%%%%%%%%%%%%%%%%%%%%%%%%%%%%%%%%%%%%%%%%%%%%%%%%%%%%%%%%%%%%%%%%%%%%%%%%%%%%%
%%%%%%%%%%%%%%%%%%%%%%%%%%%%%%%%%%%%%%%%%%%%%%%%%%%%%%%%%%%%%%%%%%%%%%%%%%%%%%%
% APPENDIX
%%%%%%%%%%%%%%%%%%%%%%%%%%%%%%%%%%%%%%%%%%%%%%%%%%%%%%%%%%%%%%%%%%%%%%%%%%%%%%%
%%%%%%%%%%%%%%%%%%%%%%%%%%%%%%%%%%%%%%%%%%%%%%%%%%%%%%%%%%%%%%%%%%%%%%%%%%%%%%%
% \newpage
% \appendix
% \onecolumn
% \section{You \emph{can} have an appendix here.}

% You can have as much text here as you want. The main body must be at most $8$
% pages long. For the final version, one more page can be added. If you want, you
% can use an appendix like this one.

% The $\mathtt{\backslash onecolumn}$ command above can be kept in place if you
% prefer a one-column appendix, or can be removed if you prefer a two-column
% appendix.  Apart from this possible change, the style (font size, spacing,
% margins, page numbering, etc.) should be kept the same as the main body.
% %%%%%%%%%%%%%%%%%%%%%%%%%%%%%%%%%%%%%%%%%%%%%%%%%%%%%%%%%%%%%%%%%%%%%%%%%%%%%%%
% %%%%%%%%%%%%%%%%%%%%%%%%%%%%%%%%%%%%%%%%%%%%%%%%%%%%%%%%%%%%%%%%%%%%%%%%%%%%%%%

\end{document}